# Moiré-driven skyrmion family

Kuan He[1], Meng-Han Li[1], Shi-Da Fan[1], Zi-Bin Lin[2], Xiao-Ying Zhuang[3], Cheng-Lin Han[1], Li-Qun Chen[4], Zhao-Dong Xu[5,6]*, Xue-Feng Zhu[2]*, Tian-Zhi Yang[1]*

(Dated: November 20, 2025)

**Affiliations**

[1]*School of Mechanical Engineering and Automation, Northeastern University, 110819 Shenyang, China*

[2]*School of Physics and Innovation Institute, Huazhong University of Science and Technology, Wuhan 430074, China*

[3]*Institute of Photonics, Faculty of Mathematics and Physics, Leibniz University Hannover, 30167, Hannover, Germany*

[4]*School of Science, Harbin Institute of Technology, Shenzhen 518055, China*

[5]*Institute of Dynamics and Smart Disaster Prevention Northeastern University, 110819 Shenyang, China*

[6]*China Pakistan Belt and Road Joint Laboratory on Smart Disaster Prevention of Major Infrastructures Southeast University Nanjing 210096 China*

***To whom all correspondence should be addressed.** xuzhaodong@mail.neu.edu.cn (Z.-D. X.); xfzhu@hust.edu.cn (X.-F. Z.); yangtianzhi@me.neu.edu.cn (T.-Z. Y.).

**Abstract**

Skyrmion family members, such as skyrmions, bimerons, and skyrmioniums, have been recently observed in quantum, solid-state, water, and magnetic systems. However, it remains challenging and crucial to identify a single platform for observing their coexistence and evolution. Here, we describe a bilayer twisted moiré elastic system as a controllable platform for the generation of skyrmion family members with distinct topological charges across different wave systems. Our experimental results further reveal that the twist angle induces a synergistic evolution between lattice symmetry and topological characteristics, enabling the mutual transformation and stable coexistence of different skyrmion family members within a single system. More importantly, we demonstrate that such a platform supports the discontinuous transport of Lamb-wave–induced topological textures, revealing phason-like dynamics within the quasiperiodic structure. This work opens a new avenue for designing all-in-one topological wave devices.

**Introduction**

Members of the skyrmion family have recently attracted widespread attention across condensed-matter physics *(1-3)*, optics *(5-9)*, acoustics *(10-12)*, and even water-wave systems *(13-15)*, owing to their topologically protected stability and distinctive dynamical behaviors. For example, optical skyrmion lattices have been observed in the evanescent electric fields of surface plasmon polaritons (SPPs) interfering within hexagonally symmetric resonators *(5)*; meron-like topological spin-angular-momentum (SAM) textures have been generated via spin-orbit interactions of light *(16)*; and skyrmion lattices have been realized in acoustic near-field evanescent waves *(11)* as well as in elastic-wave systems supporting hybrid spin states *(12)*. These nontrivial states, protected by their topological charge, exhibit remarkable robustness against defects and external perturbations, and hold significant potential for applications in information storage, energy transport, and wave manipulation. However, most existing studies have focused on the generation and control of individual skyrmion family members within a single physical system, while systematic investigations into the coupling, competition, and interconversion among different members on the same platform remain scarce. This gap hinders the evolution of topological wave physics from isolated states toward integrated multi-state architectures.

The emergence of twistronics *(17)* has offered a general paradigm for addressing this challenge. When two van-der-Waals layers are twisted by a small angle, the resulting moiré superlattice *(18)* can dramatically reshape the electronic band structure *(19)* and give rise to pronounced correlation effects *(20)*. The flat-band superconductivity observed in the so-called "magic-angle" bilayer graphene highlights the profound potential of geometric manipulation in quantum systems, inspiring the extension of twistronics concept to diverse fields such as optics, acoustics, thermotics, and fluid dynamics *(21-27)*. A variety of moiré-related unconventional phenomena have subsequently emerged, including the nearly flat dispersion observed at optical *(22)* and acoustic magic angles *(26)*; real-space nano-imaging of symmetry-broken hyperbolic phonon polaritons in monoclinic lattices *(28)*; and energy delocalization-localization transitions identified in fluidic moiré superlattices *(29)*. Collectively, these phenomena demonstrate that geometric twisting serves as a universal strategy for controlling band structures, wave localization, and topological characteristics across diverse physical systems. Similarly, twisted two lattices at specific angle break translational symmetry, giving rise to a moiré quasicrystal *(30,31)* that exhibits long-range order and higher-order rotational symmetry. The resulting distinctive band-structure *(32)* and localization characteristics *(33-35)* introduce an additional geometric degree of freedom for the emergence and manipulation of topological states. Consequently, the geometric design space enabled by moiré superlattices and moiré quasicrystals is expected to transcend the limitations of conventional physical platforms and unlock new routes toward the coexistence and controllable transformation of skyrmion family members.

In this study, we propose a moiré-induced platform for the generation of skyrmion family members. By introducing geometric twisting into a bilayer system, periodic and quasiperiodic spatial modulations are formed, enabling geometrically programmable creation of diverse skyrmion members. Due to their rich modal diversity and complex mode-coupling behavior, elastic-wave metamaterials have attracted significant attention. Their phase velocity is much lower than the speed of light, resulting in much shorter wavelengths at the same frequency and thereby enabling on-chip integration and subwavelength manipulation. *(36-43)*. In addition, it exhibits extremely low energy loss, ensuring long-distance and high-fidelity transmission, and is largely immune to environmental disturbances, guaranteeing the robustness of topological effects. These attributes make elastic waves an ideal platform for investigating the coupling and interconversion among various topological textures. In elastic plates, the interference between antisymmetric Lamb waves propagating along the upper and lower plate boundaries, together with the hybrid coupling of longitudinal and transverse components at the free surfaces, generates mixed spin interactions. Under the modulation of moiré twist angles, the system realizes stable coexistence and reversible transformation among different skyrmion family members. Furthermore, tuning the initial phases of the upper and lower Lamb waves not only induces transitions between distinct topological textures but also gives rise to topologically protected discontinuous motions that exhibit phason-like dynamical behavior. More importantly, the proposed mechanism does not depend on any specific physical field but arises from the cooperative evolution of lattice symmetry and topological characteristics (see details in Supplementary Text S5). This universality enables its extension to various wave systems, including optical, acoustic, and hydrodynamic waves, thereby opening new avenues for information processing, energy transport, and wave manipulation *(44-46)*.

**Results**

*Formation and evolution of the skyrmion family*

To elucidate the unique role of moiré twisting in controlling topological textures, we establish a universal physical platform based on a moiré bilayer elastic-wave metasurface to investigate the generation and evolution of the skyrmion family (Fig. 1A). By introducing both periodic and quasiperiodic modulations through moiré geometry, the evolution and stability of distinct topological members were systematically investigated as a function of the twist angle. This platform supports the hybridization of longitudinal and transverse waves, thereby producing Lamb waves with mixed spin characteristics. Furthermore, the interference of in-plane Lamb waves with different symmetries enables the realization of various skyrmion family members. Under the plane-strain assumption *(47, 48)*, the displacement field of antisymmetric Lamb wave propagating along the *x* direction can be expressed in terms of both out-of-plane and in-plane components

$$u_x(x,z,t) = [ikA\sin(\kappa_p z) - \kappa_s B\sin(\kappa_s z)]e^{i(kx-\omega t)}$$
$$u_z(x,z,t) = [\kappa_p A\cos(\kappa_p z) - ikB\cos(\kappa_s z)]e^{i(kx-\omega t)}, \quad (1)$$

where *A* and *B* are undetermined amplitude coefficients, $k$ and $\omega$ denote the wavenumber and angular frequency of the Lamb wave, and $\kappa_p = \sqrt{\omega^2/c_p^2 - k^2}$ and $\kappa_s = \sqrt{\omega^2/c_s^2 - k^2}$ represent the longitudinal and shear wavenumbers perpendicular to the propagation direction, respectively (see details in Supplementary Text S1). From the displacement expression in Eq. (1), the in-plane and out-of-plane components exhibit a constant phase difference of $\pi/2$, indicating that the motion trajectory of each particle in the thin plate over one vibration cycle is elliptical. The resulting elliptical polarization effectively introduces a pseudospin degree of freedom into the elastic-wave system, thereby providing the physical foundation for the formation of skyrmion family members.

For a bilayer moiré elastic-wave metasurface with a square lattice, commensurate twist angles are given by $\theta_1 = 2\tan^{-1}[p/q]$, where *p* and *q* are coprime integers, and the corresponding moiré period is $L_1 = a\sqrt{p^2 + q^2}$. For the metasurface with a triangular lattice, commensurate twist angles of the form $\theta_2 = \cos^{-1}[l^2 + 4lj + j^2/2(l^2 + lj + j^2)]$ are also considered, where *l* and *j* are coprime integers, and the moiré period is given by $L_2 = a\sqrt{l^2 + lj + j^2}$. Here, *a* denotes the lattice constant. As shown in Fig. 1B, the bilayer with a square lattice forms a moiré superlattice at $\theta_1 = 22.6°$ ($p = 1$, $q = 5$), while the bilayer with a triangular lattice produces a moiré quasicrystal at $\theta_2 = 30°$. In both lattice configurations, the excitation sources in the upper and lower layers preserve $C_4$ and $C_6$ symmetries, respectively (Fig. 1B), through the uniform distribution of eight or twelve antisymmetric Lamb-wave sources along the boundaries. This configuration ensures the spatial stability of the resulting skyrmion family members.

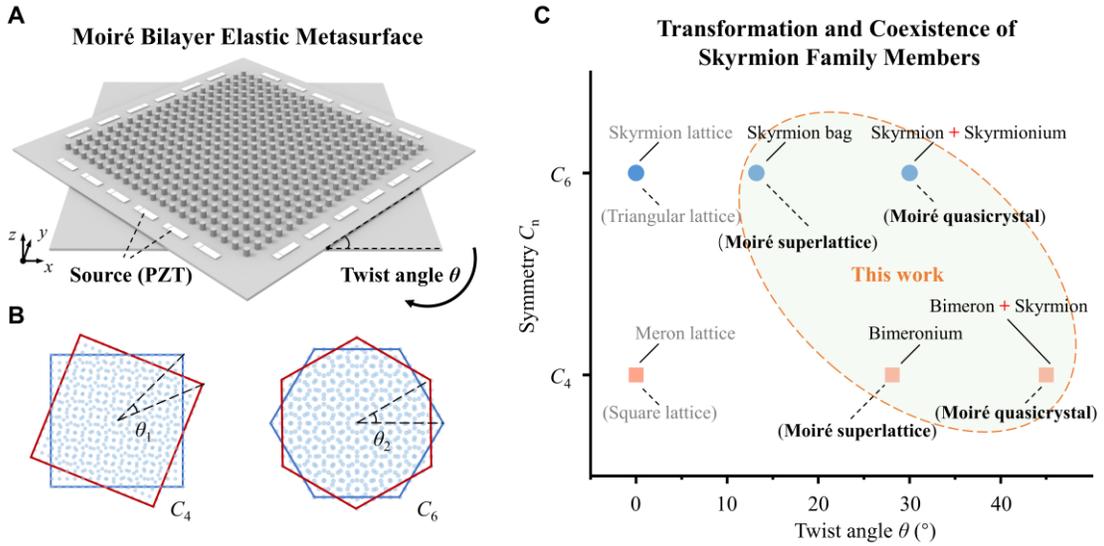

**Fig. 1. Manipulation of skyrmion family members in a moiré bilayer elastic-wave metasurface.** (**A**) Schematic of a moiré bilayer formed by stacking two identical elastic-wave metasurface plates with a relative twist angle $\theta$. Piezoelectric transducers (PZTs) are uniformly placed on the top and bottom surfaces to excite antisymmetric Lamb waves in the plates. (**B**) The left panel shows the moiré superlattice formed by two square lattices twisted by $\theta_1 = 22.6°$, whereas the right panel shows the moiré quasicrystal formed by two triangular lattices twisted by $\theta_2 = 30°$. (**C**) Adjusting the lattice symmetry and twist angle enables the generation, coexistence, and transformation of various skyrmion family members such as skyrmion lattices, meron lattices, bimeroniums, skyrmion bags, and skyrmioniums within a single platform.

*Theoretical framework for the skyrmion family*

First, a theoretical model of a moiré bilayer elastic-wave metasurface with a square-lattice arrangement was developed to investigate the skyrmion family members generated at different twist angles. In the global coordinate system $(x, y, z)$, where $z$ denotes the plate thickness direction, planar Lamb waves propagating in the antisymmetric A0 mode are considered. A plane wave propagates with an azimuthal angle $\varphi$ has wave vector $\boldsymbol{k} = k(\cos\varphi, \sin\varphi)$. In the lower plate, $n$ pairs of counter-propagating ($s = \pm 1$) planar Lamb waves are excited along azimuthal directions $\varphi_p$ ($p = 1, \cdots, n$). The displacement field formed by the interference of these $n$ pairs of waves is expressed in Eq. (2).

$$\overline{\mathbf{u}}^{(1)}(\mathbf{r}, z) = 2i \sum_{p=1}^{n} \left[ A_t^{(p)}(z)\sin(k\xi_p)\mathbf{t}(\varphi) + A_z^{(p)}(z)\cos(k\xi_p)\mathbf{z} \right], \qquad (2)$$

where $\mathbf{t}(\varphi) = (\cos\varphi, \sin\varphi)$ is the in-plane unit vector, $\mathbf{r} = (x, y)$ is the position vector, $\xi_p = \mathbf{t}(\varphi_p)\mathbf{r}$, $A_t^{(p)}(z)$ and $A_z^{(p)}(z)$ denote the amplitudes of the in-plane and out-of-plane components, respectively. The upper elastic-wave metasurface plate is uniformly rotated by an angle $\theta$ relative to the lower plate. The in-plane rotation is characterized by a two-dimensional (2D) rotation matrix (see details in Eq. (A21)), which introduces the moiré modulation effect. Consequently, the displacement field of the upper elastic-wave metasurface plate resulting from multi-wave interference is obtained, as expressed in Eq. (3).

$$\overline{\mathbf{u}}^{(2)}(\mathbf{r},z) = 2i \sum_{p=1}^{n} \left[ A_t^{(p)}(z)\sin(k\xi_p^{(\theta)})\mathbf{t}_\theta(\varphi_p) + A_z^{(p)}(z)\cos(k\xi_p^{(\theta)})\mathbf{z} \right]. \quad (3)$$

Since the amplitudes $A_t^{(p)}(z)$ and $A_z^{(p)}(z)$ are identical in the two layers, the total displacement field at a fixed depth $z = z_0$ can be expressed as

$$\overline{\mathbf{u}}(\mathbf{r}) = \overline{\mathbf{u}}^{(1)}(\mathbf{r},z_0) + \overline{\mathbf{u}}^{(2)}(\mathbf{r},z_0). \quad (4)$$

From a physical perspective, skyrmion family members constitute a class of vector textures stabilized by topological protection, whereas from a geometric perspective, they can be viewed as vector mappings of a parameter sphere onto a 2D real space. By modifying the arrangement or orientation of vectors on the parameter sphere, different types of skyrmion family members can be generated. The topological properties of these textures are characterized by the skyrmion number

$$S = \frac{1}{4\pi} \iint_\Omega \mathbf{n} \cdot (\partial_x \mathbf{n} \times \partial_y \mathbf{n}) dx dy, \quad (5)$$

where $s = \mathbf{n} \cdot (\partial_x \mathbf{n} \times \partial_y \mathbf{n})$ is the skyrmion number density, $\mathbf{n} = \frac{\overline{\mathbf{u}}}{\|\overline{\mathbf{u}}\|}$ represents the normalized 3D displacement vector field, and $\Omega$ demotes the spatial region that defines the topological textures. The skyrmion number can be further decomposed into two topological indices, $p$ and $m$, where $p$ (polarity) is defined by the orientation of the out-of-plane vector component, and $m$ (vorticity) is determined by the rotational sense of the in-plane field components in space (see details in Supplementary Text S2). In the moiré bilayer elastic-wave metasurface, modulation of the interlayer twist angle induces a pronounced evolution of the topological textures. As illustrated in Fig. 1C, the system evolves from a periodic lattice to a moiré superlattice and ultimately to a moiré quasicrystal, corresponding to the emergence of distinct skyrmion family members. Theoretical predictions and numerical simulations further reveal both the existence and evolution of these topological textures. For moiré bilayer structures with different symmetries, the corresponding 3D displacement vector fields are constructed, as shown in Figs. 2A-2F. Among these states, a bimeron can be regarded as a composite of two merons with opposite polarities, each representing half of the mapping from the parameter sphere. Therefore, merons do not occur in isolation but typically appear as periodic lattices (Fig. 2A) or as paired combinations (Fig. 2C). At a twist angle of $\theta_1 = 45°$ (Figs. 2B and 2C), skyrmions and bimerons coexist within the same configuration, whereas phase modulation drives the bimerons to be relocated to the central region (Fig. 3D). Under the combined control of twist angle and phase modulation, a bimeronium state *(49)* can likewise be realized (see details in Fig. S8).

The skyrmion bag *(4,6,50)* generated in the triangular lattice (Fig. 2E) can be regarded as a composite topological state composed of multiple skyrmions. The outer envelope exhibits a polarity opposite to that of the internal skyrmions, such that the overall topological charge can be expressed as $S_{bag} = N-1$, where $N$ denotes the number of internal skyrmions. By varying the interlayer twist angles, skyrmion bags containing different numbers of enclosed skyrmions are observed (see Supplementary Text S4). The formation mechanism of such topological textures originates from the superposition and relative rotation of two skyrmion lattices within the same plane, as illustrated at $\theta_1 = 0°$ in Fig. 2D, where the lattices in the upper and lower layers coincide. When a skyrmion in one lattice is rotated into an interstitial site between two skyrmions in the unrotated lattice, the interference effect generates a localized field distribution exhibiting a specific polarity. As the twist angle increases, the localized region gradually closes to form an envelope, eventually resulting in a skyrmion bag that encloses multiple individual skyrmions. When the interlayer twist angle reaches 30°, the translational symmetry of the system is broken, giving rise to a moiré quasicrystal exhibiting twelvefold rotational symmetry. Under this high-symmetry constraint, twelve skyrmions spontaneously arrange into a circular ring at the periphery, while the central skyrmion family transforms into a skyrmionium *(51)* composed of two nested skyrmions with opposite polarities.

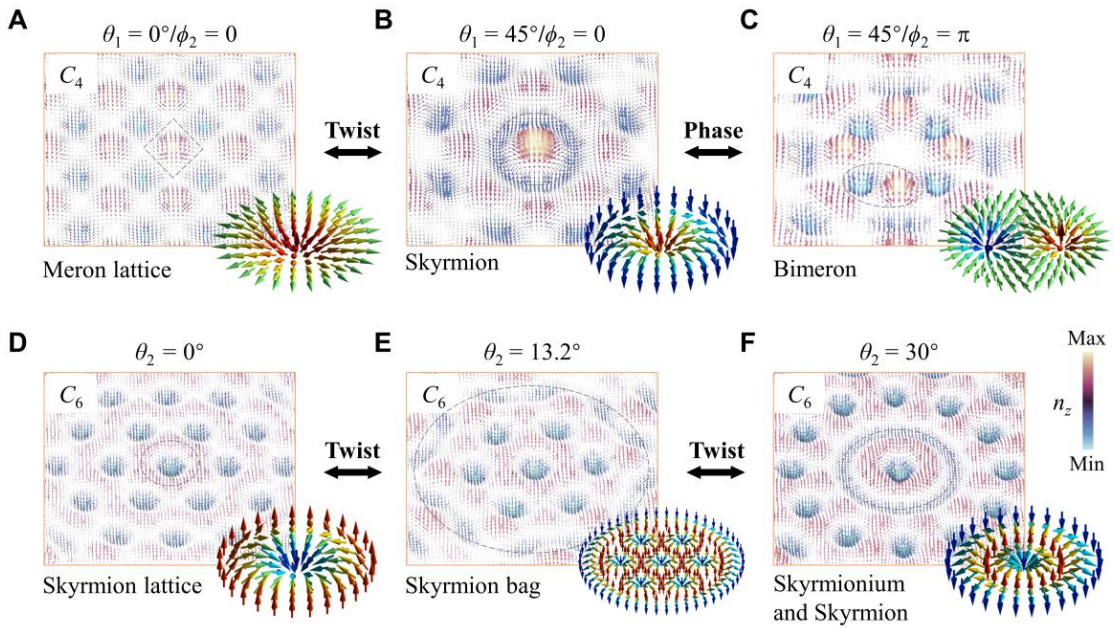

**Fig. 2. 3D displacement vector fields of distinct skyrmion family members.** (**A**) Meron lattice formed in a square-lattice configuration. (**B**) and (**C**) Moiré-induced skyrmion and bimeron states in a square-lattice bilayer twisted by $\theta_1 = 45°$, generated by upper and lower Lamb waves with initial phases ($\phi_1 = 0, \phi_2 = 0$) and ($\phi_1 = 0, \phi_2 = \pi$), respectively. (**D**) Skyrmion lattice formed in a triangular-lattice configuration. (**E**) Skyrmion-bag state emerging in a moiré superlattice formed by twisting a triangular-lattice bilayer by $\theta_2 = 13.2°$. (**F**) In a moiré quasicrystal formed by twisting a triangular-lattice bilayer by $\theta_2 = 30°$, a central skyrmionium is surrounded by a dodecagonal skyrmion-ring array. Insets in Figs. (**A**)-(**F**) show the corresponding 3D vector-field distributions of each skyrmion family member.

For the square-lattice configuration (Fig. 3A), with $n = 2$ and the azimuthal angles $\varphi_p = 0°$ and $\varphi_p = 90°$, theoretical results reveal that the system support various skyrmion family members within the square lattice, moiré superlattice, and moiré quasicrystal regimes (Fig. 3D). At $\theta_1 = 0°$, the platform exhibits a meron-lattice topological texture whose spatial distribution closely corresponds to the 3D displacement vector field shown in Fig. 2A. At $\theta_1 = 45°$, the bilayer structure adopts a moiré quasicrystalline configuration lying between crystalline and amorphous states and exhibits a hybrid topological texture composed of skyrmions and bimerons. The regions enclosed by the purple and green solid rings correspond to the 3D vector fields shown in Figs. 2B and 2C, respectively, thereby revealing the coupling and coexistence mechanisms of the two states under moiré-quasicrystalline conditions.

In addition, the excitation phase acts as an effective degree of freedom for manipulating skyrmion family members *(52)*. By tuning the relative phase difference ($\phi$) between the upper and lower Lamb waves, transformations among skyrmion family members can be induced (see Supplementary Figs. S7-S9), and discontinuous translation and positional interchange of the topological textures can be realized in a controllable manner (see Fig. 3D and Fig. S10). In the calculations shown in Fig. 3D, the initial phase of the upper-layer Lamb wave is set to $\phi_1 = 0$. At a twist angle of $\theta_1 = 45°$, when no phase difference is applied between the two layers, the central position of the topological texture remains stable, as indicated by the purple solid ring. However, when a phase difference is introduced ($\phi = \phi_2 - \phi_1 = \pi$), the topological textures enclosed by the purple and green solid rings exhibit a pronounced positional interchange, while their internal configurations remain intact owing to topological protection. Such phenomenon originates from phason-like dynamics characteristic of quasiperiodic systems *(53-55)*.

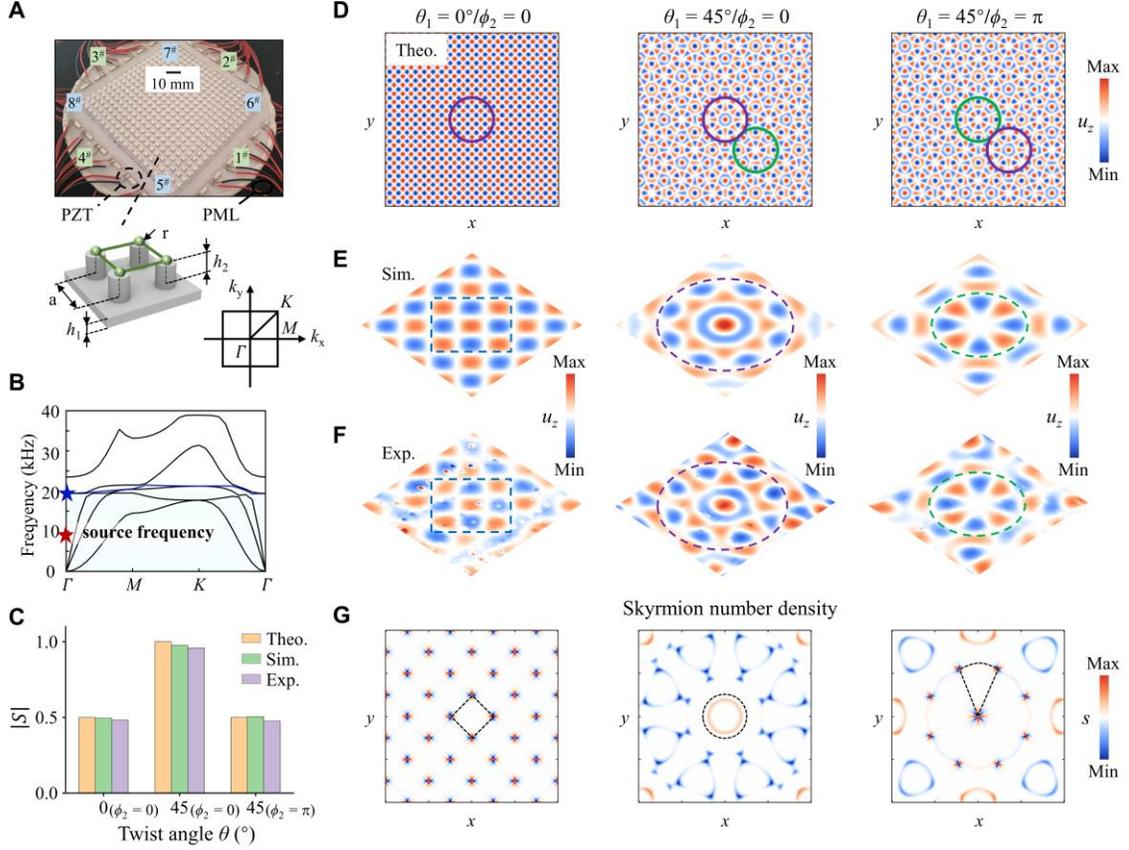

**Fig. 3. Moiré bilayer elastic metasurface with a square-lattice arrangement: theoretical prediction, experimental validation, and numerical analyses of skyrmion family members.** (**A**) Schematic of the experimental setup. The moiré bilayer metasurface is driven by PZTs numbered 1[#]~4[#] on the top layer and 5[#]~8[#] on the bottom layer. The inset illustrates the 2×2 square lattice and its corresponding first Brillouin zone. (**B**) Simulated dispersion relation of the square lattice. (**C**) Comparison of skyrmion numbers obtained from theoretical predictions, numerical simulations, and experimental measurements. (**D**)-(**F**), Out-of-plane displacement fields of skyrmion family members at different twist angles, obtained from theoretical models, numerical simulations, and experimental measurements. (**G**) Corresponding skyrmion number density for panels (**D**)-(**F**).

*Experimental observation of skyrmion family*

To validate the theoretical predictions, numerical simulations and experimental measurements were subsequently performed. PZTs were uniformly arranged along the edges of the top and bottom layers to excite antisymmetric Lamb waves propagating in the ±$k$ directions within the plate plane. The simulated out-of-plane displacement field is shown in Fig. 3E, and the corresponding in-plane components ($|u_\parallel| = \sqrt{u_x^2 + u_y^2}$) are presented in Fig. S6. The resulting 3D displacement vector fields for the square-lattice configuration agree well with those shown in Figs. 2A-C. The dispersion relation of the square lattice was also calculated, as shown in Fig. 3B. Below the first cutoff frequency, the desired topological textures can in principle be excited at arbitary frequency *(12)*, and an excitation frequency of $f$ = 8900 Hz was selected for the experiments. The dynamic displacement field of the sample surface, was then scanned and recorded using a laser Doppler vibrometer (PSV-500), as presented in Fig. 3F. The measured displacement field shows excellent agreement with the theoretical predictions within the solid-ring region indicated in Fig. 3D.

Based on the 3D displacement fields obtained from theoretical calculations, numerical simulations, and experimental measurements, the skyrmion number density associated with different skyrmion family members was evaluated (Fig. 2G). The skyrmion numbers corresponding to the different topological configurations were determined by integrating the skyrmion number density over the regions enclosed by dashed lines in Fig. 3C. The results demonstrate strong consistency in the topological invariants obtained from theoretical derivations, numerical simulations, and experimental measurements, thereby verifying the coexistence and transformation of skyrmion family members under different twist angles and confirming the robustness and stability of these topological textures.

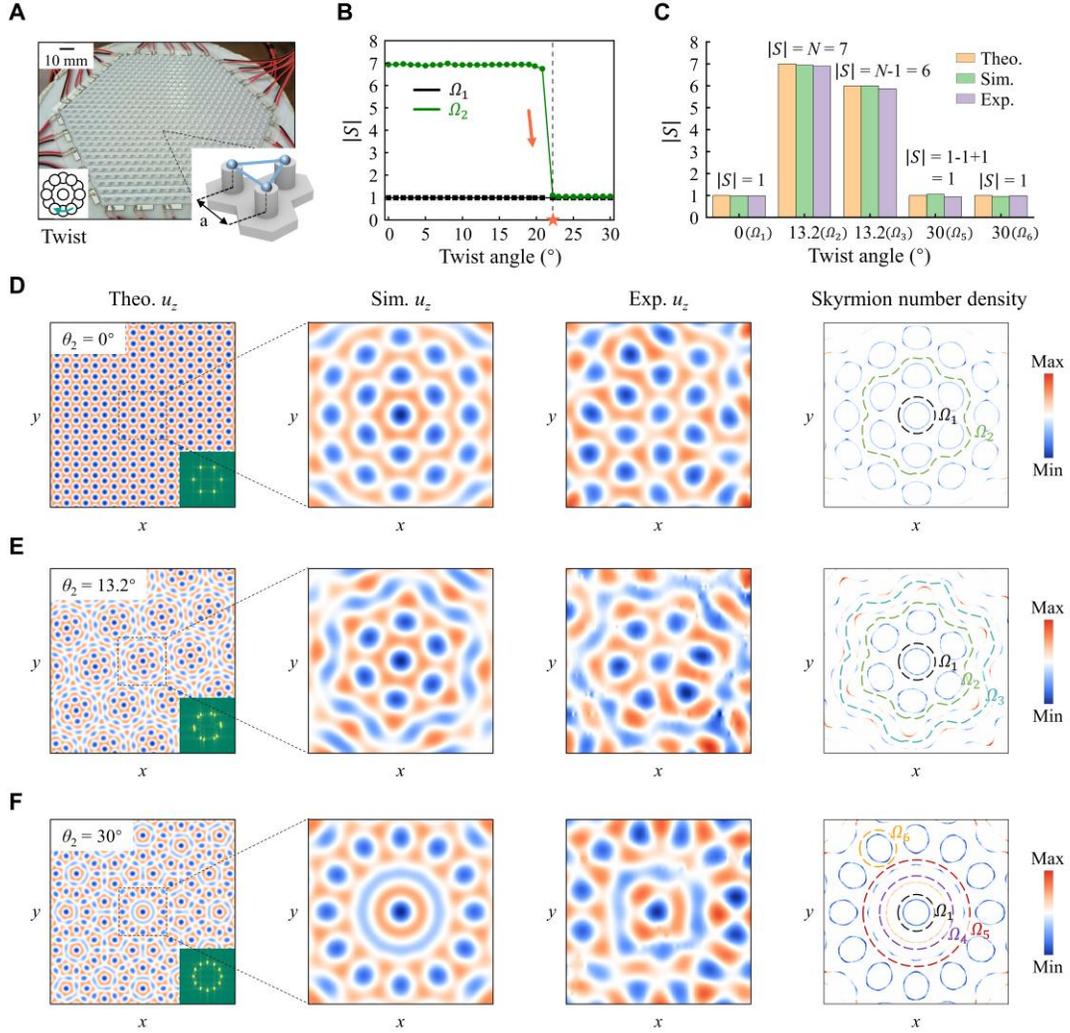

**Fig. 4. Moiré bilayer elastic metasurface with a triangular-lattice arrangement: theoretical, experimental, and numerical comparisons of the evolution of skyrmion family members.** (**A**) Experimental setup of the moiré bilayer metasurface with a triangular-lattice arrangement. The inset illustrates the schematic of the triangular lattice. (**B**) Variation of the skyrmion number within two integration regions ($\Omega_1$ and $\Omega_2$) as a function of the interlayer twist angle $\theta_2$. (**C**) Comparison of skyrmion numbers obtained from theoretical models, numerical simulations, and experimental measurements. (**D**)-(**F**), Skyrmion family members observed at different twist angles: a moiré lattice at $\theta_2 = 0°$, a moiré superlattice at $\theta_2 = 13.2°$, and a moiré quasicrystal at $\theta_2 = 30°$. For each case, the columns present the out-of-plane displacement fields obtained from theoretical calculations, numerical simulations, and experimental measurements, followed by the corresponding skyrmion- number density.

To further verify the universality of the proposed twisting mechanism, we extend the lattice configuration from a square to a triangular arrangement and construct a bilayer moiré elastic-wave metasurface based on a triangular lattice (Fig. 4A). The triangular lattice constant was kept identical to that of the square lattice (*a*), which enables a direct comparison of topological responses under different symmetries on the same geometric scale. Fig. 4B shows the variation of the skyrmion number as a function of the interlayer twist angle. These integration regions correspond to the green and black rings marked in Fig. 4D. The results indicate that, for all twist angles considered, region $\Omega_1$ remains occupied by a single skyrmion. In contrast, within region $\Omega_2$, when $\theta_2 < \theta_{2*} = 20.8°$, the region contains seven individual skyrmions, whereas for $\theta_2 > \theta_{2*}$, the skyrmion number decreases to approximately one, which corresponds to the formation of a skyrmionium. The corresponding topological charge satisfies $S = 1-1+1 = 1$, indicating that the outer annular domain wall and the inner skyrmion reorganize into a concentric structure and thereby drive the transition from multiple skyrmions to a skyrmionium state.

Theoretical predictions, numerical simulations, and experimental measurements under different twist angles are presented in Figs. 4D-F. All three show strong consistency in both the spatial distribution of the topological textures and the corresponding skyrmion numbers. The inset of Fig.4D further shows the 2D fast Fourier transform of the displacement field, clearly revealing the evolution of symmetry from a periodic lattice to a moiré superlattice and finally to a moiré quasicrystal. Specifically, at $\theta_2 = 0°$, the system forms a typical Néel-type skyrmion lattice with $m = 1$ (hedgehog configuration). At $\theta_2 = 13.2°$ ($l = 2, j = 3$), a skyrmion bag appears, in which several skyrmions are enclosed by an outer envelope of opposite polarity. When $\theta_2 = 30°$, a moiré-quasicrystalline pattern emerges, with the central structure evolving into a skyrmionium and the periphery self-organizing into a dodecagonal skyrmion ring array. The skyrmion numbers were calculated for the seven central skyrmions at $\theta_2 = 0°$ and the twelve peripheral skyrmions at $\theta_2 = 30°$, both yielding values close to unity (see Supplementary Fig. S12 for details). Furthermore, the skyrmion numbers of the skyrmion family members at different twist angles were evaluated (Fig. 4C), which show strong consistency among theoretical predictions, numerical simulations, and experimental measurements. These findings validate the theoretical model and demonstrate that the proposed moiré bilayer platform enables topological transitions and coexistence across different lattice symmetries, underscoring the universality and broad applicability.

**Discussion**

In summary, we propose and experimentally validates a physical platform for the moiré-induced generation of the skyrmion family. By tuning the relative twist angle between the two metasurface layers, the platform enables the cooperative evolution of lattice symmetry and topological properties, thereby providing a unified description of the emergence and transformation of multiple topological textures. Using an elastic-wave metasurface as the experimental platform, we observed the generation, coexistence, and transformation of skyrmion family members with distinct topological charges, such as skyrmions, bimerons, skyrmion lattices, skyrmioniums, meron lattices, bimeroniums, and skyrmion bags. The observed behavior demonstrates the robustness of these topological textures. Further phase modulation enables the discontinuous transport of topological textures while preserving their structural integrity, revealing the phason-like dynamical characteristics inherent to quasiperiodic systems. This work establishes a moiré topological evolution framework applicable to diverse wave systems, offering both theoretical and experimental insights into the mechanisms and control of topological texture formation. Moreover, the framework offers new design strategies and technical routes for topological particle manipulation *(56,57)* as well as on-chip topological devices.

**Materials and Methods**

*Experimental setup and measurements*

The bilayer moiré elastic-wave metasurface samples were fabricated by 3D printing with epoxy resin. A suspension configuration was employed for vibration-mode measurements to minimize boundary effects. The substrate plate has a thickness of $h_1 = 2$ mm, and the cylindrical resonators have a radius of $r = 2$ mm and a height of $h_2 = 5$ mm. Both square and triangular lattice configurations were implemented, with a lattice constant of $a = 10$ mm. The material parameters are: density $\rho = 1160$ kg/m$^3$, Poisson's ratio $v = 0.42$, and Young's modulus $E = 2.62$ GPa. Antisymmetric Lamb waves were excited by an array of lead zirconate titanate (PZT-5H) patches bonded to the top and bottom layers and driven via a laser Doppler vibrometer controller. The out-of-plane displacement fields over all measurement points on the sample surface were reconstructed using a PSV-500-1D scanning head. The experimental data were post-processed in MATLAB, where spatial interpolation was applied to enhance spatial resolution. To ensure measurement reliability and accuracy, each point was measured five times, and the average value was used for analysis.

*Numerical simulations*

Numerical simulations of planar Lamb-wave propagation in the bilayer moiré elastic-wave metasurface were conducted using the Solid Mechanics module of COMSOL Multiphysics 6.3. The structural and material parameters in the simulations were kept identical to those used in the experiments. To simulate the incidence of planar Lamb waves, acceleration boundary conditions were applied to the upper and lower plates, and perfectly matched layers were placed around the bilayer structure to suppress boundary reflections. To ensure the accuracy and convergence of the finite-element calculations, the maximum mesh element size was chosen as one twelfth of the shortest wavelength considered. Frequency-domain simulations produced the 3D displacement fields of the bilayer moiré metasurface at different twist angles. The Eigenfrequency module was also employed to calculate the dispersion relations within the first Brillouin zone of a unit cell. The numerical results were further post-processed in MATLAB to obtain the corresponding Fourier spectra and skyrmion- number density, from which the skyrmion numbers were subsequently evaluated

**Acknowledgments**

**Funding:** This work is supported by the National Natural Science Foundation of China under grant numbers 12232014 (T.-Z.Y.), Fundamental Research Funds for the Central Universities and the Ten Thousand Talents under grant numbers N2403027 (T.-Z.Y.), China Scholarship Council under grant numbers 202406080122 (C.-L.H.), Fundamental Research Funds for the Central Universities under grant numbers N2403003 (C.-L.H.), Science Center Program of National Natural Science Foundation of China under grant numbers 62188101 (L.-Q. C.), National Natural Science Foundation of China under grant numbers 12132002 (L.-Q. C.), National Key Research and Development Program of China under grant numbers 2020YFA0211400 (X.-F. Z.) and 2020YFA0211401 (X.-F. Z.)

**Author contributions:**
    Conceptualization: X.-F.Z., Z.-D.X., X.-Y. Z., L.-Q. C., T.-Z.Y.
    Methodology: K.H., M.-H. L., C.-L.H., M.-H.L.
    Validation: K.H., S.-D.F., Z.-B. L.
    Visualization: K.H.
    Supervision: L.-Q. C., T.-Z.Y.
    Writing—original draft: K.H.
    Writing—review & editing: X.-F.Z., Z.-D.X., T.-Z.Y.

**Competing interests:**

The authors declare no competing interests.

**Data and materials availability:**

All data are available in the main text or the supplementary materials. Any additional requests for information can be directed to and will be fulfilled by the corresponding authors.

39. R. Zhu, X. N. Liu, G. K. Hu, C. Sun, and G. Huang, Negative refraction of elastic waves at the deep-subwavelength scale in a single-phase metamaterial. *Nat. Commun.* 5, 5510 (2014).

40. M. Jiang, Y.-F. Wang, B. Assouar, Y.-S. Wang, Scattering-free modulation of elastic shear-horizontal waves based on interface-impedance theory. *Phys. Rev. Appl.* 20, 054020 (2023).

41. S. Wang, Z. Hu, Q. Wu, H. Chen, E. Prodan, R. Zhu, G. Huang, Smart patterning for topological pumping of elastic surface waves. *Sci. Adv*. 9, eadh4310 (2023).

42. Y.-K. Ma, W. Guo, Y.-M. Cui, Y.-F. Wang, V. Laude, Y.-S. Wang, Attenuation of Lamb waves in coupled-resonator viscoelastic waveguide. *Int. J. Mech. Sci.* 285, 109790 (2025).

43. S. An, T. Liu, L. Cao, Z. Gu, H. Fan, Y. Zeng, L. Cheng, J. Zhu, B. Assouar, Multibranch elastic bound states in the continuum. *Phys. Rev. Lett*. 132, 187202 (2024).

44. N. Nagaosa and Y. Tokura, Topological properties and dynamics of magnetic skyrmions. *Nat. Nanotechnol.* 8, 899–911 (2013).

45. L. Zhao, C. Hua, C. Song, W. Yu, W. Jiang, Realization of skyrmion shift register. *Sci. Bull*. 69, 2370–2378 (2024).

46. A. Fert, N. Reyren, and V. Cros, Magnetic skyrmions: Advances in physics and potential applications. *Nat. Rev. Mater.* 2, 17031 (2017).

47. H. Lamb, On waves in an elastic plate. *Proc. R. Soc. Lond. A* 93, 114–128 (1917).

48. J. L. Rose, *Ultrasonic Guided Waves in Solid Media.* (Cambridge Univ. Press, Cambridge, 2014).

49. X. Zhang, J. Xia, M. Ezawa, O. A. Tretiakov, H. T. Diep, G. Zhao, X. Liu, Y. Zhou, A frustrated bimeronium: static structure and dynamics. *Appl. Phys. Lett*. 118, 052411 (2021).

50. D. Foster, C. Kind, P. J. Ackerman, J. B. Taylor, M. R. Dennis, and I. I. Smalyukh, Two-dimensional skyrmion bags in liquid crystals and ferromagnets. *Nat. Phys.* 15, 655–659 (2019).

51. A. G. Kolesnikov, M. E. Stebliy, A. S. Samardak, and A. V. Ognev, Skyrmionium—high velocity without the skyrmion Hall effect. *Sci. Rep.* 8, 16966 (2018).

52. S. Hayami, T. Okubo, and Y. Motome, Phase shift in skyrmion crystals. *Nat. Commun.* 12, 6927 (2021).

53. M. Lin, X. Gou, Z. Xie, A. Yang, L. Du, X. Yuan, Photonic quasicrystal of spin angular momentum. *Sci. Adv*. 11, eadv3938 (2025).

54. T. A. Corcovilos and J. Mittal, Two-dimensional optical quasicrystal potentials for ultracold atom experiments. *Appl. Opt.* 58, 2256–2263 (2019).

55. B. Freedman, G. Bartal, M. Segev, R. Lifshitz, D. N. Christodoulides, and J. W. Fleischer, Wave and defect dynamics in nonlinear photonic quasicrystals. *Nature* 440, 1166–1169 (2006).